\documentclass[conference]{IEEEtran}
\IEEEoverridecommandlockouts
\usepackage{cite}
\usepackage{citesort}
\usepackage{graphicx}
\usepackage{amsmath}
\usepackage{amsfonts}
\usepackage{xcolor}
\usepackage{amssymb,bm,upgreek}
\usepackage{algorithm}
\usepackage{algorithmic}
\usepackage{multirow}
\usepackage{transparent}
\usepackage{verbatim}
\usepackage{subfigure}

\usepackage{import}

\DeclareMathOperator\erf{erf}
\DeclareMathOperator*{\E}{\mathbb{E}}

\newcommand{\prob}{\textnormal{Pr}}
\newcommand{\agg}{\textnormal{agg}}
\newcommand{\s}{\textnormal{s}}
\newcommand{\mole}{\textnormal{molecule}}

\newcommand{\m}{\textnormal{m}}
\newcommand{\self}{\textnormal{self}}
\newcommand{\coop}{\textnormal{c}}

\newtheorem{remark}{Remark}
\newtheorem{theorem}{Theorem}

\def\BibTeX{{\rm B\kern-.05em{\sc i\kern-.025em b}\kern-.08em
    T\kern-.1667em\lower.7ex\hbox{E}\kern-.125emX}}

\begin{document}

\title{Expected Density of Cooperative Bacteria\\in a 2D Quorum Sensing Based\\Molecular Communication System}

\author{\IEEEauthorblockN{Yuting Fang${}^\dag$, Adam Noel${}^\ddag$, Andrew W. Eckford${}^\sharp$, and Nan Yang${}^\dag$}
\IEEEauthorblockA{${}^\dag$Research School of Electrical, Energy and Materials Engineering, Australian National University, Canberra, ACT, Australia\\}
\IEEEauthorblockA{${}^\ddag$School of Engineering, University of Warwick, Coventry, UK\\}
\IEEEauthorblockA{${}^\sharp$Department of Electrical Engineering and Computer Science, York University, Toronto, ON, Canada}}

\maketitle

\begin{abstract}
The exchange of small molecular signals within microbial populations is generally referred to as quorum sensing (QS). QS is ubiquitous in nature and enables microorganisms to respond to fluctuations in living environments by working together. In this study, a QS-based molecular communication system within a microbial population in a two-dimensional (2D) environment is analytically modeled. Microorganisms are randomly distributed on a 2D circle where each one releases molecules at random times. The number of molecules observed at each randomly-distributed bacterium is first derived by characterizing the diffusion and degradation of signaling molecules within the population. Using the derived result and some approximation, the expected density of cooperative bacteria is derived. Our model captures the basic features of QS. The analytical results for noisy signal propagation agree with simulation results where the Brownian motion of molecules is simulated by a particle-based method. Therefore, we anticipate that our model can be used to predict the density of cooperative bacteria in a variety of QS-coordinated activities, e.g., biofilm formation and antibiotic resistance.
\end{abstract}

\IEEEpeerreviewmaketitle

\section{Introduction}\label{sec:intro}

Quorum sensing (QS) is a ubiquitous approach for microbial communities to respond to a variety of situations in which monitoring the local population density is beneficial. In QS, bacteria assess the number of other bacteria they can interact with by releasing and recapturing a molecular signal in their environment. This is because a higher density of bacteria leads to more molecules that can be detected before they diffuse away. If the number of molecules detected exceeds a threshold, bacteria express target genes for a cooperative response. QS enables coordination within large groups of cells, potentially increasing the efficiency of processes that require a large population of cells working together.

Microscopic populations utilize QS to complete many collaborative activities, such as virulence and bioluminescence. These tasks play a crucial role in bacterial infections, environmental remediation, and wastewater treatment \cite{7506074}. Since the QS process is highly dependent on signaling molecules, the accurate characterization of releasing, diffusion, degradation, and reception of such molecules across the environment in which bacteria live is very important to understand and control QS. This can help us to prevent undesirable bacterial infections and lead to new environmental remediation methods \cite{0006655}.

There are growing research efforts to study the coordination of bacteria via QS. \cite{0006655,WEST2007,Lindsay2017WhenIP} investigated the cooperative behavior of bacteria using simulation or biological experiments. \cite{8278046,7181698,7935509,8422668,7397847,7742378} mathematically modeled bacterial behavior coordination, but relied on abstract or simplifying models to represent the molecular diffusion channel for the purpose of understanding how behavior evolves over time.

It is of significant theoretical and practical importance to develop accurate models of QS communication systems, particularly in terms of signal propagation and responsive behaviors. We address this problem in the present paper using the knowledge of QS, mass diffusion, stochastic geometry, and probability processes. We for the first time analytically model a QS-based molecular communication (MC) system by characterizing the diffusion and degradation of signaling molecules, considering bacteria that are randomly spatially distributed on a bounded circle where each one continuously emits molecules at random times. Unlike most existing MC studies that consider a one-dimensional or three-dimensional (3D) environment, we consider a 2D environment since a 2D environment facilitates future experimental validation of our current theoretical work. Biological experiments, especially with bacteria, are usually conducted in a 2D environment, e.g., bacteria residing on a petri dish and the formation of biofilms \cite{biofilm}.

Our model captures the basic features of QS by adopting the following assumptions:
\begin{enumerate}
\item We consider bacteria that are randomly spatially distributed on a bounded circle since the location of bacteria cannot be manually controlled in reality.
\item We consider bacteria that continuously emit  molecules at random times since the sporadic molecule emission process is stochastic in practice.
\item We consider that each bacterium is both a transmitter (TX) and a receiver (RX) which captures the features of emission and reception of molecules at bacteria.
\item We adopt the same decision strategy at bacteria as QS, i.e., the concentration threshold-based strategy.
\item Our model accounts for the random propagation of signaling molecules based on reaction-diffusion equations.
\end{enumerate}
In consideration of these realistic assumptions, we make the following contributions:
\begin{enumerate}
\item We analytically derive the asymptotic channel response (i.e., the expected number of molecules observed) at a circular RX due to continuous emission of molecules at a) one point TX and b) randomly-distributed point TXs on a circle in a 2D environment. Using this result, we can determine with high accuracy the concentration observed by each organism in a QS environment.
\item We obtain a model for cooperative behavior in QS by deriving an approximate expression for the expected density of cooperative bacteria, using the asymptotic channel response at each bacterium.
\end{enumerate}

To demonstrate the benefits of our contribution, we validate the accuracy of our analytical results via a particle-based simulation method where we track the random walk of each signaling molecule over time. We note that the asymptotic channel response can generally be applied to any context where a TX is continuously releasing molecules into a 2D environment. We also note that our results could be readily extended to a 3D environment by changing the 2D area integrations to 3D volume integrations. Importantly, our model can be used to predict with high accuracy the effect of diffusion and chemical molecular reactions on the concentration of molecules observed by bacteria and the expected density of responsive cooperators, since our model takes into consideration the random motion of molecules based on reaction-diffusion equations.

We use the following notations: $|\vec{x}|$ denotes Euclidean norm of a vector $\vec{x}$. $\overline{N}$ denotes the mean of a random variable (RV) $N$ and $\E\{\overline{N}\}$ denotes the expectation of $\overline{N}$ over a spatial random point process. $K_n(Z)$ denotes modified $n$th order Bessel function of the second kind.

\section{System Model}\label{sec:system model}

\begin{figure}[!t]
    \centering
    \includegraphics[width=0.75\columnwidth]{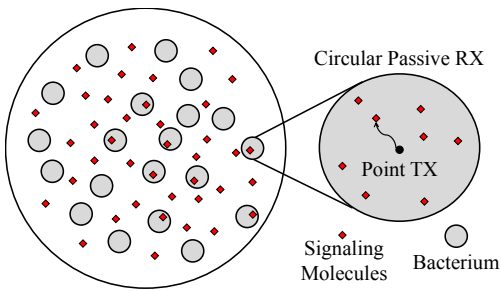}
    \vspace{0mm}
    \caption{A population of bacteria randomly distributed on a circle according to a 2D spatial point process, where each bacterium acts as a point TX and as a circular passive RX. The molecules diffuse into and out of the bacteria.}\label{fig:model}
\end{figure}

We consider an unbounded two-dimensional (2D) environment. A population of bacteria is spatially distributed on a bounded circle $S_1$ with radius $R_1$ centered at $(0,0)$ according to a 2D point process with constant density $\lambda$, as shown in Fig.~\ref{fig:model}. Point processes are commonly used to model randomly-distributed locations of bacterial populations, e.g., \cite{Jeanson1493}. We denote $\vec{x_i}$ as the location of the center of the $i$th bacterium. We denote $\Phi\left(\lambda\right)$ as the random set of bacteria locations. We consider bacteria behavior analogous to QS, i.e., 1) emit signaling molecules; 2) detect the concentration of signaling molecules; and 3) decide to cooperate if the concentration exceeds a threshold. In the following, we detail the emission, propagation, and reception of signaling molecules, and decision-making by the bacteria.

\begin{figure}[!t]
    \centering
    \includegraphics[width=0.75\columnwidth]{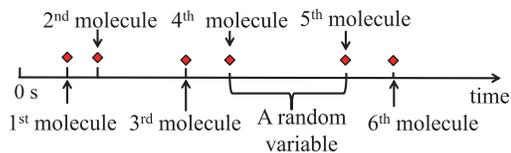}
    \caption{An example of release times due to continuous emission of molecules at a bacterium according to a random process.}\label{fig:emit}
\end{figure}

\textbf{Emission}: We model bacteria as point TXs. The $i$th bacterium continuously emits $A$ molecules from $\vec{x_i}$ at random times according to an independent random process with constant rate $q$ ${\mole}/{\s}$, as shown in Fig. \ref{fig:emit}. We note that continuous emission does not mean there is no time interval between two successive emissions of molecules. Instead, we model the time interval as a random variable and the expected interval length is inversely proportional to the emission rate.

\textbf{Propagation}: All $A$ molecules diffuse independently with a constant diffusion coefficient $D$. They can degrade into a form that cannot be detected by the bacteria, i.e., $A\overset{k}\rightarrow\emptyset$, where $k$ is the reaction rate constant in $\s^{-1}$. If $k=0$, this degradation is negligible. Since we consider a single type of molecules, we only mention ``the molecules'', instead of ``$A$  molecules'', in the remainder of this paper.

\begin{figure}[!t]
    \centering
    \includegraphics[height=1.1in]{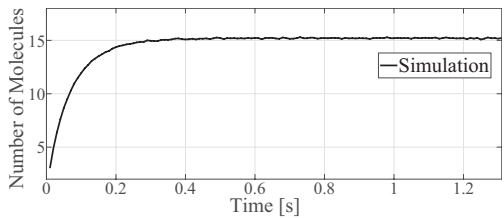}
    \caption{The time-varying expected number of molecules observed $\overline{N}_{\agg}^{\dag}\left(\vec{x_i},t|\lambda\right)$ versus time $t$ for $R_1 = 20\,{\mu}\m$, $\lambda=7.9\times10^{-2}/{\mu}\m^2$, and $\vec{x_i}=(10,10)$. For other simulation details, please see Sec. \ref{sec:Numerical}.}\label{radiusAndtime}
\end{figure}

\textbf{Reception}: We model the $i$th bacterium as a circular passive receiver (RX) with radius $R_0$ and area $S_0$ centered at $\vec{x_i}$. Bacteria perfectly count molecules if they are within $S_0$. Since the molecules released from all bacteria may be observed by the $i$th bacterium, the number of molecules observed at the $i$th bacterium at time $t$, ${N}_{\agg}^{\dag}\left(\vec{x_i},t|\lambda\right)$, is given by $N_{\agg}^{\dag}\left(\vec{x_i},t|\lambda\right)=\sum_{\vec{x_j}\in\Phi\left(\lambda\right)}N\left(\vec{x_i},t|\vec{x_j}\right)$, where $N\left(\vec{x_i},t|\vec{x_j}\right)$ is the number of molecules observed at the $i$th bacterium at time $t$ due to the $j$th bacterium. The means of $N_{\agg}^{\dag}\left(\vec{x_i},t|\lambda\right)$ and $N\left(\vec{x_i},t|\vec{x_j}\right)$ are denoted by $\overline{N}_{\agg}^{\dag}\left(\vec{x_i},t|\lambda\right)$ and $\overline{N}\left(\vec{x_i},t|\vec{x_j}\right)$, respectively. We assume that the \emph{expected} number of molecules observed at the $i$th bacterium is constant after some time. To demonstrate the suitability of this assumption, see Fig. \ref{radiusAndtime} (and an analytical proof in Sec. \ref{ObsPlayer}). In Fig. \ref{radiusAndtime}, $\overline{N}_{\agg}^{\dag}\left(\vec{x_i},t|\lambda\right)$ is independent of $t$ after time $t=0.5\,\s$. We denote time $t^{\star}_i$ as the time after which $\overline{N}_{\agg}^{\dag}\left(\vec{x_i},t|\lambda\right)$ is constant, i.e.,
\begin{align}\label{asymp}
\overline{N}_{\agg}^{\dag}\!\left(\vec{x_i},t|\lambda\right)\!|_{t>t^{\star}_i}\!=\!\lim_{t\rightarrow\infty}\!\overline{N}_{\agg}^{\dag}\!\left(\vec{x_i},t|\lambda\right)\!=\!\overline{N}_{\agg}^{\dag}\!\left(\vec{x_i},\infty|\lambda\right).
\end{align}

\textbf{Decision-Making}: We assume that the $i$th bacterium makes one decision at some time $t_i$, $t_i>t^{\star}_i$, when the expected number of observed molecules becomes stable. This assumption is reasonable since $t^{\star}_i$ is very small, e.g., $t=0.5\,\s$ in Fig. \ref{radiusAndtime}, and bacteria can reach the steady state very quickly, especially relative to the timescale of gene regulation to coordinate behavior\footnote{Based on \cite{Danino2010,Trovato2014,Surette7046}, the cooperation of bacteria is observed after the signaling molecules diffuse for at least tens of minutes.}. Also, bacteria can wait until there are enough molecules to trigger behavior change. Therefore, bacteria do not need to explicitly know whether the steady state has been reached and the precise synchronization of emission and detection is not needed. Inspired by a threshold-based strategy in QS, we assume that the $i$th bacterium decides as follows: $B\left(\vec{x_i}\right)=1$ if $N_{\agg}^{\dag}\left(\vec{x_i},\infty|\lambda\right)>\eta$; otherwise, $B\left(\vec{x_i}\right)=0$. Here, $B\left(\vec{x_i}\right)$ is the decision of the $i$th bacterium, ``1'' denotes cooperation, ``0'' denotes non-cooperation, and $\eta$ is a decision threshold. For compactness, we remove $\infty$ in all notation in the remainder of this paper since we assume that bacteria use asymptotic observations under noisy signal propagation to make decisions.

We acknowledge the major simplifications that we make for the tractability of our analysis, as follows:
\begin{enumerate}
\item We consider bacteria do not move. This assumption is appropriate when bacteria swim very slowly or for some non-motile bacteria, e.g., coliform and streptococci.
\item We assume that bacteria are passive RXs that do not interact with molecules. This is because that the observations at multiple bacteria are \emph{correlated} if RXs interact with molecules, which makes analysis much cumbersome.
\item Each bacterium makes one decision based on one sample of the observed signal. Modeling evolutionary or repeat behavior coordination with noisy signal propagation is interesting for future work, as identified in \cite{Game_tutorial}, although \cite{Game_tutorial} did not model evolutionary behavior.
\item We simplify bacteria as a point source emitting molecules isotropically into the environment. Considering the emission of molecules from imperfect TXs into particular directions is left for future work.
\item We assume that the average emission rate of molecules is constant. This assumption is appropriate for scenarios where bacteria transit from being selfish to ramping up molecule production before the emission rate has changed.
\end{enumerate}
We emphasize that our system model still captures the basic features of QS and considers noisy signal propagation among a field of bacteria, although the aforementioned simplifications are made.

\section{2D Channel Response}\label{sec:2D}

In this section, we derive the channel response, i.e., the expected number of molecules observed at RX, due to continuous emission of molecules from TX(s), in the following cases: 1) a point TX and 2) randomly distributed TXs. These analyses lay the foundations for our derivations of the observations at bacteria and expected density of cooperators in Sec. \ref{sec:game}.

To derive the channel response due to continuous emission, we first review the channel response due to one impulse emission. Based on \cite[eq.~(3.4)]{Crank_book} and the fact that the molecule degradation introduces a decaying exponential term as in \cite[eq.~(10)]{6712164}, the channel response $C\left(\vec{r},\tau\right)$ at the point defined by $\vec{r}$ at the time $\tau$ due to an impulse emission of one molecule from the point at $(0,0)$ at time $\tau=0$ into an unbounded 2D environment, is given by
\begin{align}\label{impulse}
C\left(\vec{r},\tau\right) = \frac{1}{\left(4\pi D \tau\right)}\exp\left(-\frac{|\vec{r}|^{2}}{4D\tau}-k\tau\right).
\end{align}

We next derive the asymptotic channel response based on \eqref{impulse} and we assume that the RX is a circular passive observer $S_0$ centered at $\vec{b}$ with radius $R_0$.

\subsection{One Point TX}

In this subsection, we present the asymptotic channel response due to one point TX. We also present the special case when the TX is at the center of the RX, since each bacterium receives the molecules released from not only other bacteria but also itself. We finally simplify the channel response using the uniform concentration assumption (UCA) \cite{Noel2013}.

\subsubsection{Arbitrary $\vec{b}$}

The asymptotic channel response can be obtained by multiplying $C\left(\vec{r},\tau\right)$ by the emission rate $q$, integrating over $S_0$, and then integrating over all time to infinity. By doing so, the asymptotic channel response at the circular RX with radius $R_0$ centered at $\vec{b}$, due to continuous emission at random times with rate $q$ from the point $(0,0)$ since time $t=0$, is given by\footnote{In this paper, arbitrary $\vec{b}$ includes the special case $|\vec{b}|=0$.}
\begin{align}\label{point-circle}
\overline{N}(\vec{b}) = &\;\int_{\tau=0}^{\infty}\int_{r=0}^{R_0}\int_{\theta=0}^{2\pi}qC\left(\vec{r_1},\tau\right)rd\theta drd\tau,\nonumber\\
=&\;\int_{\tau=0}^{\infty}\int_{r=0}^{R_0}\int_{\theta=0}^{2\pi}\frac{q}{(4\pi D \tau)}\nonumber\\
&\times\exp\left(-\frac{|\vec{r_1}|^{2}}{4D\tau}-k\tau\right)rd\theta drd\tau,\nonumber\\
=&\;\int_{\tau=0}^{\infty}\int_{r=0}^{R_0}\int_{\theta=0}^{2\pi}\frac{q}{(4\pi D \tau)}\nonumber\\
&\hspace{-5mm}\times\exp\left(-\frac{{|\vec{b}|}^2+r^2+2{|\vec{b}|}r\cos\theta}{4D\tau}-k\tau\right)rd\theta drd\tau,
\end{align}
where $\vec{r_1}$ is a vector from $(0,0)$ to a point within the RX.

\subsubsection{$|\vec{b}|=0$}

We have the following theorem:
\begin{theorem}\label{theo,point-circle}
The asymptotic channel response at the circular RX with radius $R_0$, due to continuous emission with rate $q$ from the center of this RX since time $t=0$, is given by
\begin{align}\label{point-circle,self}
\overline{N}_{\self}=&\;\lim_{|\vec{b}|\rightarrow0}\overline{N}(\vec{b})\nonumber\\
= &\;\frac{q}{k}\bigg(1-{\frac{\sqrt{k}R_0}{\sqrt{D}}K_1\bigg(\sqrt{\frac{k}{D}}R_0\bigg)}\bigg).
\end{align}
\begin{IEEEproof}
The proof is given in Appendix \ref{app}.
\end{IEEEproof}
\end{theorem}

\subsubsection{UCA}

We simplify \eqref{point-circle} by assuming that the concentration of molecules throughout the circular RX is uniform and equal to that at the center of the RX. This assumption is accurate if $|\vec{b}|$ is relatively large and thus it is inaccurate when $|\vec{b}|=0$. Using this assumption, we rewrite \eqref{point-circle} as
\begin{align}\label{point}
\overline{N}(\vec{b}) = &\;\pi R_0^2\int_{\tau=0}^{\infty}qC\left(\vec{b},\tau\right)d\tau.
\end{align}

We then employ \cite[eq.~(3.471)]{gradshteyn2007} to solve \eqref{point} as
\begin{align}\label{point1}
\overline{N}(\vec{b}) =&\;\frac{q R_0^2}{2D}K_0\left(|\vec{b}|\sqrt{\frac{k}{D}}\right).
\end{align}

\subsection{Randomly Distributed TXs}

In this subsection, we consider that \emph{many} point TXs are randomly distributed on a circle $S_1$ according to a point process with density $\lambda$. The circle $S_1$ is centered at $(0,0)$ with radius $R_1$. We represent $\vec{a}$ as the location of an arbitrary point TX $a$ and the random set of TXs' locations is denoted by $\Phi\left(\lambda\right)$. We denote the channel response at the RX at time $t$ due to TX $a$ by $\overline{N}\left(\vec{b},t|\vec{a}\right)$ and the aggregate channel response at the RX at time $t$ due to all TXs by $\overline{N}_{\agg}\left(\vec{b},t|\lambda\right)=\sum_{\vec{a}\in\Phi\left(\lambda\right)}\overline{N}\left(\vec{b},t|\vec{a}\right)$. We denote $\E\left\{\overline{N}_{\agg}\left(\vec{b}|\lambda\right)\right\}$ as the expected $\overline{N}_{\agg}\left(\vec{b},t|\lambda\right)$ over the point process. We next derive $\overline{N}_{\agg}\left(\vec{b},t|\lambda\right)$ and then simplify it using the UCA. We have the following theorem:
\begin{theorem}[Arbitrary $\vec{b}$]\label{theo,circle-cirle, agg}
The expected aggregate asymptotic channel response at the circular RX with radius $R_0$ centered at $\vec{b}$, due to continuous emission with rate $q$ since time $t=0$ from randomly distributed TXs on circle $S_1$ with density $\lambda$, is given by
\begin{align}
&\;\E\left\{\overline{N}_{\agg}\left(\vec{b}|\lambda\right)\right\}\nonumber\\
=&\;\E\Bigg\{\sum_{\vec{a}\in\Phi\left(\lambda\right)}\overline{N}\left(\vec{b}|\vec{a}\right)\Bigg\}\nonumber\\
=&\;\int_{|\vec{r}|=0}^{R_1}\int_{\varphi=0}^{2\pi}\int_{\tau=0}^{\infty}\int_{|\vec{r_0}|=0}^{R_0}\int_{\theta=0}^{2\pi}\frac{q\exp\left(-\frac{\Upsilon(\vec{b})^2}{4D\tau}-k\tau\right)}{(4\pi D \tau)}\nonumber\\
&\times|\vec{r_0}|\lambda |\vec{r}|\,d\theta\,d|\vec{r_0}|\,d\tau\,d\varphi\,d|\vec{r}|\label{circle-cirle, aggv1}\\
= &\;\lambda \int_{|\vec{r}|=0}^{R_1}\int_{\varphi=0}^{2\pi}\int_{|\vec{r_0}|=0}^{R_0}\int_{\theta=0}^{2\pi}K_0\left(\sqrt{\frac{k}{D}} \Upsilon(\vec{b})\right)\nonumber\\
&\times \frac{q}{2D\pi}|\vec{r_0}||\vec{r}|\,d\theta\,d|\vec{r_0}| \,d\varphi\,d|\vec{r}|\label{circle-cirle, agg},
\end{align}
where
\begin{align}\label{A}
\Upsilon(\vec{b}) = \sqrt{\Omega(\vec{b})+|\vec{r_0}|^2+2\sqrt{\Omega(\vec{b})}|\vec{r_0}|\cos\theta},
\end{align}
and $\Omega(\vec{b})={|\vec{b}|}^2+|\vec{r}|^2+2{|\vec{b}|}|\vec{r}|\cos\varphi$.
\begin{IEEEproof}
The proof of Theorem \ref{theo,circle-cirle, agg} is omitted here due to space limitation. It can be proven using Campbell's theorem \cite{Haenggi:2012:SGW:2480878}, \eqref{point-circle}, and the law of cosines.
\end{IEEEproof}
\end{theorem}

Under the UCA, we use \eqref{point} to approximate the expectation of $\overline{N}_{\agg}\left(\vec{b}|\lambda\right)$ as
\begin{align}\label{circle-cirle, agg2}
\E\left\{\overline{N}_{\agg}\left(\vec{b}|\lambda\right)\right\} \approx &\;\int_{|\vec{r}|=0}^{R_1}\int_{\varphi=0}^{2\pi}\frac{q{R_0}^2}{2D}\lambda |\vec{r}|\,d\varphi\,d|\vec{r}|\nonumber\\
&\times K_0\left(\sqrt{\frac{k}{D}\Omega(\vec{b})}\right).
\end{align}
The numerical results in Sec. \ref{sec:Numerical} will demonstrate the accuracy of the approximation of the UCA in \eqref{point1} and \eqref{circle-cirle, agg2}. We note that time-varying channel responses are also of interest. Thus, we discuss them in the following remark:
\begin{remark}\label{re:time-varying}
It can be shown that the \emph{time-varying} channel response $\overline{N}\left(\vec{b},t\right)$ and $\E\left\{\overline{N}_{\agg}\left(\vec{b},t|\lambda\right)\right\}$ can be obtained by replacing $\infty$ with $t$ in \eqref{point-circle} and \eqref{circle-cirle, aggv1}, respectively. 	
\end{remark}

\section{Analysis of Density of Bacterial Cooperators}\label{sec:game}

In this section, we aim to evaluate the expected density of cooperators over the point process $\E\left\{\overline{\lambda_{\coop}}\right\}$, where $\lambda_{\coop}$ denotes the density of cooperators. To this end, we first analyze the expected aggregate asymptotic number of observed molecules at the $i$th bacterium, $\overline{N}_{\agg}^{\dag}\left(\vec{x_i}|\lambda\right)$, for a given realization of the point process.

\subsection{Observation by Bacteria}\label{ObsPlayer}
We recall that the $i$th bacterium observes molecules in the environment released from all bacteria (also including the molecules released from itself). Thus, we have
\begin{align}\label{obsExp}
\overline{N}_{\agg}^{\dag}\left(\vec{x_i}|\lambda\right)=&\;\sum_{\vec{x_j}\in\Phi\left(\lambda\right)}\overline{N}\left(\vec{x_i}|\vec{x_j}\right)\nonumber\\
=&\;\overline{N}\left(\vec{x_i}|\vec{x_i}\right)+\sum_{\vec{x_j}\in\Phi\left(\lambda\right)/\vec{x_i}}\overline{N}\left(\vec{x_i}|\vec{x_j}\right),
\end{align}
where $\overline{N}\left(\vec{x_i}|\vec{x_i}\right)=\overline{N}_{\self}$ and $\overline{N}_{\self}$ is given in \eqref{point-circle,self}. We then approximate the second term of the second line in \eqref{obsExp} as
\begin{align}\label{obsExp2}
\sum_{\vec{x_j}\in\Phi\left(\lambda\right)/\vec{x_i}}\overline{N}\left(\vec{x_i}|\vec{x_j}\right)\approx &\;\E\Bigg\{\sum_{\vec{a}\in\Phi\left(\acute{\lambda}\right)}\overline{N}\left(\vec{x_i}|\vec{a}\right)\Bigg\},
\end{align}
where $\acute{\lambda}={\left(\lambda\pi R_1^2-1\right)}/{\pi R_1^2}$. In \eqref{obsExp2}, we use the \emph{expected} channel response \emph{over the point process} to approximate the channel response under \emph{one} realization of this point process. Also, we consider a new density $\acute{\lambda}$ to keep the average number of bacteria the same after the approximation of \eqref{obsExp2}. Our numerical results in Sec. \ref{sec:Numerical} will confirm the accuracy of the approximation of \eqref{obsExp2}. We further re-write \eqref{obsExp2} as
\begin{align}\label{obsExp2,1}
\E\Bigg\{\sum_{\vec{a}\in\Phi\left(\acute{\lambda}\right)}\overline{N}\left(\vec{x_i}|\vec{a}\right)\Bigg\}= &\;\E\left\{\overline{N}_{\agg}\left(\vec{x_i}|\acute{\lambda}\right)\right\},
\end{align}
where $\E\left\{\overline{N}_{\agg}\left(\vec{x_i}|\acute{\lambda}\right)\right\}$ can be evaluated by replacing $|\vec{b}|$ and $\lambda$ with $|\vec{x_i}|$ and $\acute{\lambda}$, respectively, in \eqref{circle-cirle, agg} or \eqref{circle-cirle, agg2}.

\begin{remark}\label{re:asymptotic}
We have analytically found that $\overline{N}_{\agg}^{\dag}\left(\vec{x_i}|\lambda\right)$ converges as time $t\rightarrow\infty$, since $\overline{N}_{\agg}^{\dag}\left(\vec{x_i}|\lambda\right)$ can be obtained via \eqref{circle-cirle, agg} (or \eqref{circle-cirle, agg2}) and \eqref{point-circle,self} and they all converge with time. This analytically proves that our assumption adopted for \textbf{Reception} in Sec. \ref{sec:system model} is valid, i.e., $\overline{N}_{\agg}^{\dag}\left(\vec{x_i}|\lambda\right)$ does not vary with time $t$ after some time.
\end{remark}

\subsection{Density of Cooperators}

In this subsection, we aim to evaluate $\E\left\{\overline{\lambda_{\coop}}\right\}$. To this end, we first analyze the binary decision at the $i$th bacterium, $B\left(\vec{x_i}\right)$, and its mean $\overline{B}_m\left(\vec{x_i}\right)$. Since $B\left(\vec{x_i}\right)$ is a Bernoulli RV, we evaluate $\overline{B}\left(\vec{x_i}\right)$ as
\begin{align}\label{prob}
\overline{B}\left(\vec{x_i}\right)=\prob\left(B\left(\vec{x_i}\right)=1\right)= 1-\prob\left(N_{\agg}^{\dag}\left(\vec{x_i}|\lambda\right)<\eta\right).
\end{align}

We recall that $N_{\agg}^{\dag}\left(\vec{x_i}|\lambda\right)$ is the sum of $N\left(\vec{x_i}|\vec{x_j}\right)$ over $j$. We note that $N\left(\vec{x_i}|\vec{x_j}\right)$ is the sum of the number of molecules observed at the $i$th bacterium at time $t=t_i$ released from the $j$th bacterium since $t=0\,\s$. Thus, the observations at the $i$th bacterium due to continuous emission at the $j$th bacterium are not identically distributed since they are released at different times. Therefore, $N\left(\vec{x_i}|\vec{x_j}\right)$ is a Poisson binomial RV since each molecule behaves independently and has a different probability of being observed at $t=t^{\star}_i$ by the $i$th bacterium due to different releasing times. Since $N_{\agg}^{\dag}\left(\vec{x_i}|\lambda\right)$ is the sum of $N\left(\vec{x_i}|\vec{x_j}\right)$, $N_{\agg}^{\dag}\left(\vec{x_i}|\lambda\right)$ is also a Poisson binomial RV. We note that modeling $N_{\agg}^{\dag}\left(\vec{x_i}|\lambda\right)$ as a Poisson binomial RV makes the evaluation of the cumulative density function (CDF) of $N_{\agg}^{\dag}\left(\vec{x_i}|\lambda\right)$ in \eqref{prob} very cumbersome, since we need to account for each probability of each molecule being observed at the $i$th bacterium released from all bacteria since $t=0\,\s$. Fortunately, using the central limit theorem \cite{ROSS201489}, we can accurately approximate $N_{\agg}^{\dag}\left(\vec{x_i}|\lambda\right)$ as a Gaussian RV. We further approximate the variance of $N_{\agg}^{\dag}\left(\vec{x_i}|\lambda\right)$ by its mean $\overline{N}_{\agg}^{\dag}\left(\vec{x_i}|\lambda\right)$. By doing so and using the CDF of a Gaussian RV \cite{ROSS201489}, we obtain
\begin{align}\label{prob1}
\overline{B}\left(\vec{x_i}\right)=&\;\prob\left(B\left(\vec{x_i}\right)=1\right)\nonumber\\
=&\;1-\frac{1}{2}\left[1+\erf\left(\frac{\eta-0.5-\overline{N}_{\agg}^{\dag}\left(\vec{x_i}|\lambda\right)}
{\sqrt{2\overline{N}_{\agg}^{\dag}\left(\vec{x_i}|\lambda\right)}}\right)\right],
\end{align}
where $\overline{N}_{\agg}^{\dag}\left(\vec{x_i}|\lambda\right)$ is evaluated in \eqref{obsExp}.
We next analyze the expected number of cooperators. We denote the number of cooperators and its mean for a given realization of the spatial point process by $Z$ and $\overline{Z}$, respectively. Since $Z=\sum_{\vec{x_i}\in\Phi\left(\lambda\right)}B\left(\vec{x_i}\right)$, we have $\overline{Z}=\sum_{\vec{x_i}\in\Phi\left(\lambda\right)}\overline{B}_m\left(\vec{x_i}\right)$. Using the Campbell's theorem \cite{Haenggi:2012:SGW:2480878}, we calculate the expected number of cooperators over the random point process as
\begin{align}\label{coop}
\E\left\{\overline{Z}\right\}= &\;\E\Bigg\{\sum_{\vec{x_i}\in\Phi\left(\lambda\right)}\overline{B}\left(\vec{x_i}\right)\Bigg\}\nonumber\\
= &\; 2\pi\lambda\int_{|\vec{r_1}|=0}^{R_1}\overline{B}\left(\vec{r_1}\right)|\vec{r_1}|d|\vec{r_1}|,
\end{align}
where $\overline{B}\left(\vec{r_1}\right)$ can be obtained by replacing $\vec{x_i}$ with $\vec{r_1}$ in \eqref{prob1}. Combining \eqref{coop}, \eqref{prob1}, \eqref{obsExp}, \eqref{circle-cirle, agg} (or \eqref{circle-cirle, agg2}), and \eqref{point-circle,self}, we rewrite $\E\left\{\overline{Z}\right\}$ as
\begin{align}\label{coop,final}
\E\left\{\overline{Z}\right\}=&\;\int_{|\vec{r_1}|=0}^{R_1}\Bigg\{-\frac{1}{2}\Bigg[1+\erf\Bigg(\frac{\eta-0.5-\overline{N}_{\agg}^{\dag}(\vec{r_1}|\lambda)}{\sqrt{2\overline{N}_{\agg}^{\dag}(\vec{r_1}|\lambda)}}\Bigg)\Bigg]\nonumber\\
&+1\Bigg\}\lambda 2\pi|\vec{r_1}|\,d|\vec{r_1}|,
\end{align}
where
\begin{align}\label{obsExp3}
\overline{N}_{\agg}^{\dag}(\vec{r_1}|\lambda)= &\;\overline{N}_{\agg}(\vec{r_1}|\acute{\lambda})+\overline{N}_{\self},
\end{align}
and $\overline{N}_{\agg}(\vec{r_1}|\acute{\lambda})$ can be obtained by substituting $|\vec{b}|$ and $\lambda$ with $|\vec{r_1}|$ and $\acute{\lambda}$, respectively, in \eqref{circle-cirle, agg} or \eqref{circle-cirle, agg2}. We finally obtain $\E\left\{\overline{\lambda_{\coop}}\right\}$ by $\E\left\{\overline{\lambda_{\coop}}\right\}=\E\left\{\overline{Z}\right\}/\pi R_1^2$.

\section{Numerical Results and Simulations}\label{sec:Numerical}

\begin{table}[!t]
\renewcommand{\arraystretch}{1.2}
\centering
\caption{Environmental Parameters}\label{tab:table1}
\begin{tabular}{c||c|c}
\hline
\bfseries Parameter &  \bfseries Symbol&  \bfseries Value \\
\hline\hline
Radius of observer & $R_{0}$ & $1\,{\mu}\m$ \\\hline
Diffusion coefficient & $D$ & $10^{-9}{\m^{2}}/{\s}$\\\hline
Emission rate & $q$ & $1 \times10^{3}{\mole}/{\s}$\\\hline
Reaction rate constant & $k$ & $1\times10^{1}/{\s}$\\\hline
\end{tabular}
\end{table}

In this section, we present simulation and numerical results to assess the accuracy of our derived analytical results and reveal the impact of environmental parameters on the number of molecules observed and density of cooperators.

\begin{figure}[!t]
    \centering
    \includegraphics[height=2.27in]{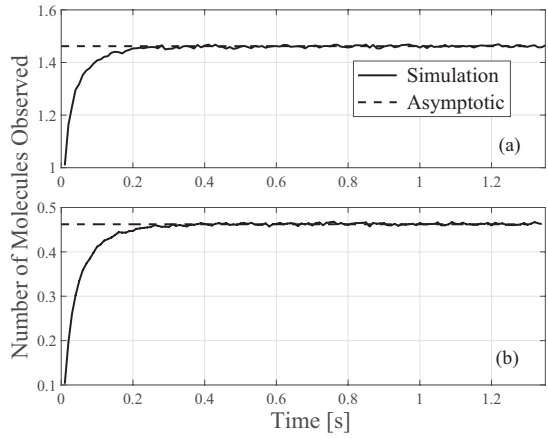}
    \caption{The expected number of molecules observed at the RX $\overline{N}\left(\vec{b},t\right)$ due to continuous emission at one TX located at $(0,0)$ versus time when the RX is located at (a) $(0,0)$ and (b) $(5\,{\mu}\m,0)$.}\label{oneRX}
\end{figure}

\begin{figure}[!t]
    \centering
    \includegraphics[height=2.27in]{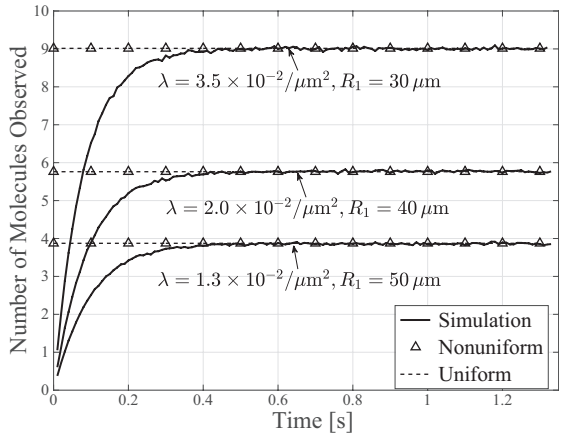}
    \caption{The expected number of molecules observed at the RX $\E\left\{\overline{N}_{\agg}\left(\vec{b},t|\lambda\right)\right\}$ due to continuous emission at randomly-distributed TXs for different environmental radius $R_1$. The average number of TXs is 100 and the RX's location is $(10\,{\mu}\m,10\,{\mu}\m)$.}\label{mulTXs}
\end{figure}

\begin{figure}[!t]
    \centering
    \includegraphics[height=2.27in]{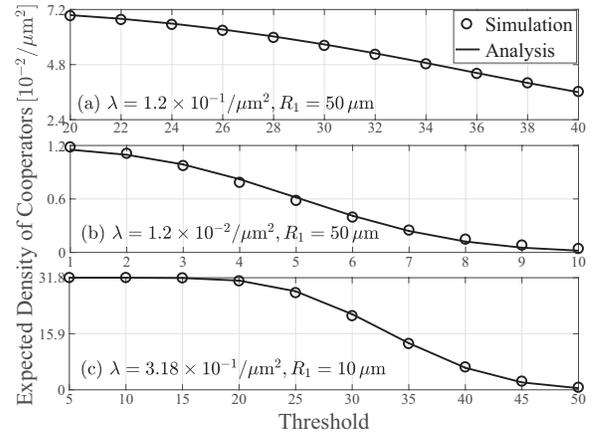}
    \caption{The expected density of cooperators over spatial PPP $\E\left\{\overline{\lambda_{\coop}}\right\}$ versus threshold $\eta$ for different population radius $R_1$ and density $\lambda$.}\label{threshold}
\end{figure}

The simulation details are as follows: The simulation environment is unbounded. We vary density $\lambda$ and bacteria community radius $R_1$. We list other fixed environmental parameters in Table \ref{tab:table1}. The value of the parameters $R_{0}$, $D$, $\lambda$, and $R_1$ are chosen to be on the same orders of those used in \cite{Danino2010,Trovato2014,Surette7046}. We simulate the Brownian motion of molecules using a particle-based method as described in \cite{Andrew2004}. The molecules are initialized at the center of bacteria. The location of each molecule is updated every time step $\Delta t$, where diffusion along each dimension is simulated by generating a normal RV with variance $2D\Delta t$. Every molecule has a chance of degrading in every time step with the probability $\exp(-k\Delta t)$. In simulations, the locations of bacteria are distributed according to a 2D Poisson point process (PPP). Each bacterium releases molecules according to an independent Poisson process, thus the times between the release of consecutive molecules at different bacteria are simulated as i.i.d exponential RVs. In Figs. \ref{oneRX}--\ref{threshold}, the simulation is repeated $10^4$ times. In Fig. \ref{oneRX}, there is one TX at a fixed location and for each realization we randomly generate molecule release times at the TX. In Figs. \ref{mulTXs} and \ref{threshold}, for each realization we randomly generate both the locations and molecule release times for all TXs (bacteria).

In Fig. \ref{oneRX}, we plot the expected number of molecules observed at the RX due to continuous emission by one TX in two cases: a) the TX is at the center of the RX and b) the distance between the TX and the RX is $5\,{\mu}\m$. In Fig. \ref{mulTXs}, we plot the expected number of molecules observed at the RX due to continuous emission by a circular field of TXs for different environmental radii and we keep the average number of bacteria fixed at 100. The asymptotic curves in Fig. \ref{oneRX}(a) and Fig. \ref{oneRX}(b) are evaluated by \eqref{point-circle,self} and \eqref{point1}, respectively. The asymptotic curves with UCA and without UCA in Fig. \ref{mulTXs} are obtained via \eqref{circle-cirle, agg2} and \eqref{circle-cirle, agg}, respectively. We first note that the expected number of molecules observed in Figs. \ref{oneRX} and \ref{mulTXs} first increases as the time increases and then becomes stable after time $t\approx0.5\,\s$. Second, we note that all asymptotic curves agree with the simulations, thereby validating the accuracy of \eqref{point-circle,self}, \eqref{point1}, \eqref{circle-cirle, agg2}, and \eqref{circle-cirle, agg}. Third, in Fig. \ref{mulTXs}, we note that the asymptotic curves with UCA and without UCA almost overlap with each other. This demonstrates the accuracy of the UCA in the derivation of the channel response where a circular field of TXs continuously emit molecules. Finally, we note that when $R_1$ decreases, the expected number of molecules increases. This is not surprising since the density of TXs is higher when $R_1$ is smaller.

In Fig. \ref{threshold}, we plot the expected density of cooperators versus threshold for different population radii and densities. The analytical curves are obtained by \eqref{coop,final} via \eqref{circle-cirle, agg2}. We see that the simulations have good agreement with the analytical curves, thereby validating the accuracy of \eqref{coop,final} and \eqref{circle-cirle, agg2}. We also see that the expected density of cooperators decreases when the threshold increases, because the probability of being cooperative at bacteria is smaller when the threshold is higher.

\begin{figure}[!t]
    \centering
    \includegraphics[height=2.27in]{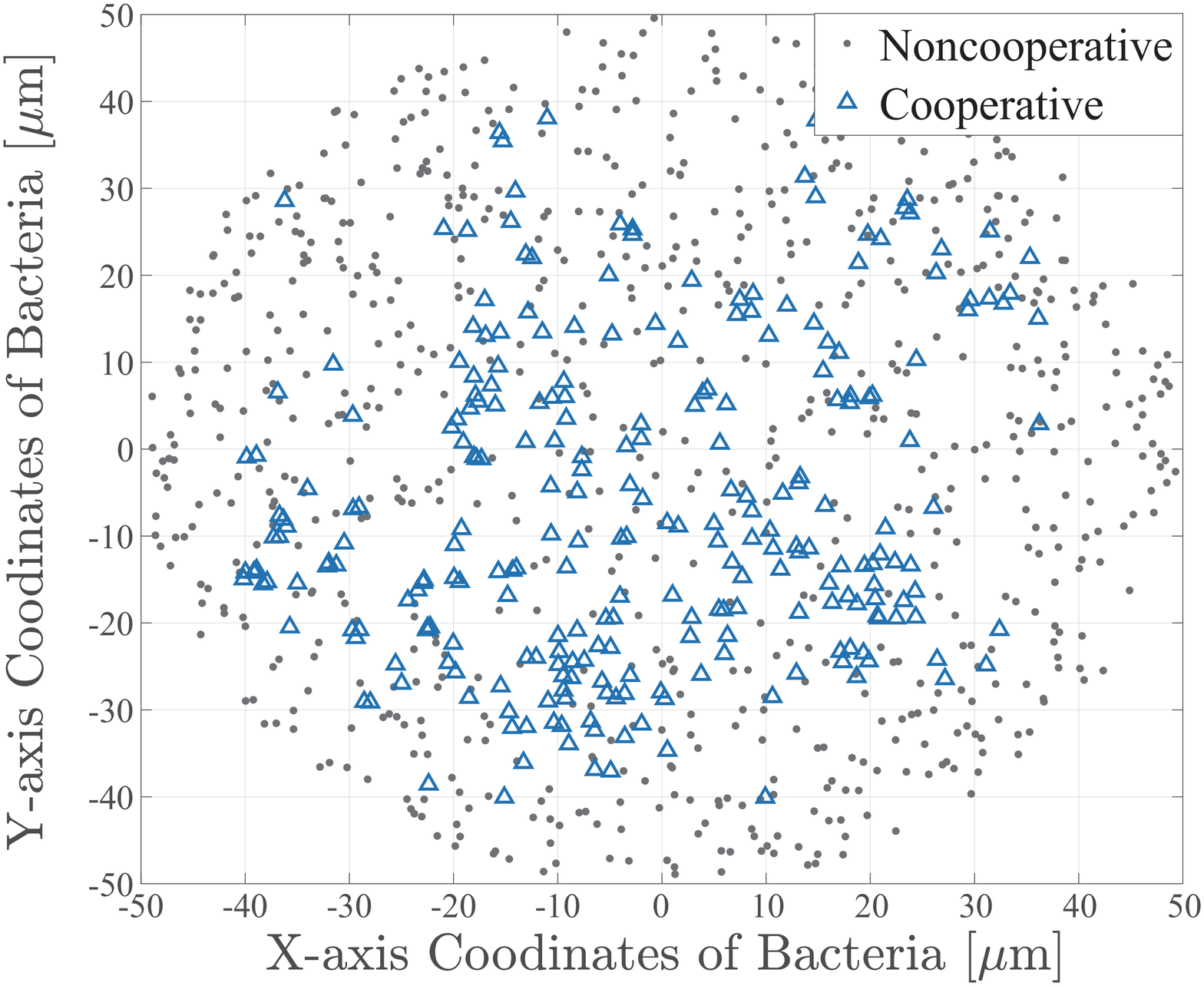}
    \caption{The spatial distribution of cooperators under one realization of randomly-distributed locations of bacteria and random molecule release times at all bacteria in a simulation with $R_1 = 50\,{\mu}\m$ and $\lambda=1.2\times10^{-1}/{\mu}\m^2$.}\label{spatial}
\end{figure}

In Fig. \ref{spatial}, we simulate the decisions of bacteria under one realization of randomly-distributed bacteria locations and random molecule release times at all bacteria. We plot the spatial distribution of cooperators in this realization. The number of cooperators around the population center is larger than that at the population edge. This is because the expected number of molecules observed at the bacteria closer to the center is higher than those observed at the bacteria located further from the center.

\section{Conclusions}\label{sec:con}

In this work, we analytically modeled a QS-based MC system of a microbial population in a 2D environment. Microorganisms were randomly distributed on a circle with a constant density where each one releases molecules at random times and with a fixed emission rate. To analyze the observations and responsive behaviors at bacteria, we first analytically derived the asymptotic channel response at a circular RX due to continuous emission of molecules at 1) one point TX and 2) randomly-distributed point TXs on a circle in a 2D environment. From this analysis, the number of molecules observed at each randomly-distributed bacterium was analyzed and the expected density of cooperative bacteria over a spatial random point process was derived. Our analytical results were validated using a particle-based simulation method.

Interesting future work includes: 1) Experimental validation of our current analytical results, 2) applying game theory to our current model with elaborated payoffs and strategies, 3) predicting the higher order statistics of the density of cooperative bacteria, and 4) relaxing simplified assumptions, e.g., non-motile bacteria and constant emission rate of molecules.

\appendices

\section{Proof of Theorem \ref{theo,point-circle}}\label{app}
Applying $|\vec{b}|=0$ to \eqref{point-circle}, we first write $\overline{N}_{\self}$ as
\begin{align}\label{point-circleAp}
\overline{N}_{\self}=&\;\int_{\tau=0}^{\infty}\int_{r=0}^{R_0}\int_{\theta=0}^{2\pi r}\frac{q \exp\left(-\frac{r^2}{4D\tau}-k\tau\right)}{(4\pi D \tau)}\,d\theta \,dr\,d\tau,\nonumber\\
=&\;\int_{\tau=0}^{\infty}\int_{r=0}^{R_0}\frac{q r}{(2D \tau)}\exp\left(-\frac{r^2}{4D\tau}-k\tau\right) \,dr\,d\tau.
\end{align}

We then apply \cite[eq.~(2.33.12)]{gradshteyn2007} given by
\begin{align}\label{exp-exp}
\int x^{m}\exp\left(-\beta x^{n}\right)dx =&\;-(\gamma-1)!\frac{\exp\left(-\beta x^{n}\right)}{n}\nonumber\\
&\times\left(\sum_{k=0}^{\gamma-1}\frac{x^{n k}}{k!\beta^{\gamma-k}}\right),
\end{align}
where $\gamma=\frac{m+1}{n}$, to \eqref{point-circleAp} and use some basic integral manipulations to rewrite \eqref{point-circleAp} as
\begin{align}\label{point-circle,self2}
\overline{N}_{\self} =&\;\int_{\tau=0}^{\infty}q\exp(-k\tau)\left(1-\exp\left(\frac{-R_0^2}{4D\tau}\right)\right)\,d\tau.
\end{align}

We finally apply \cite[eq.~(3.310)]{gradshteyn2007} given by $\int_{=0}^{\infty}\exp(-px)dx = {1}/{p}$
and \cite[eq.~(3.324.1)]{gradshteyn2007} given by
\begin{align}\label{exp2}
\int_{0}^{\infty} \exp\left(-\frac{\beta}{x}-\gamma x\right)dx = \frac{\beta}{\gamma}K_{1}(\sqrt{\beta\gamma}),
\end{align}
to \eqref{point-circle,self2} to arrive at \eqref{point-circle,self}.


\end{document}